\documentclass[sigconf]{acmart}
\usepackage{booktabs}

\AtBeginDocument{%
  \providecommand\BibTeX{{%
    \normalfont B\kern-0.5em{\scshape i\kern-0.25em b}\kern-0.8em\TeX}}}

\copyrightyear{2022}
\acmYear{2022}
\setcopyright{acmcopyright}\acmConference[EASE 2022]{The International Conference on Evaluation and Assessment in Software Engineering 2022}{June 13--15, 2022}{Gothenburg, Sweden}
\acmBooktitle{The International Conference on Evaluation and Assessment in Software Engineering 2022 (EASE 2022), June 13--15, 2022, Gothenburg, Sweden}
\acmPrice{15.00}
\acmDOI{10.1145/3530019.3530040}
\acmISBN{978-1-4503-9613-4/22/06}

\begin{document}

\title{A Hierarchical DBSCAN Method for Extracting Microservices from Monolithic Applications}


\author{Khaled Sellami}
\affiliation{%
  \institution{Laval University}
  \city{Quebec}
  \state{QC}
  \country{Canada}}
\email{khaled.sellami.1@ulaval.ca}

\author{Mohamed Aymen Saied}
\affiliation{%
  \institution{Laval University}
  \city{Quebec}
  \state{QC}
  \country{Canada}}
\email{mohamed-aymen.saied@ift.ulaval.ca}

\author{Ali Ouni}
\affiliation{%
  \institution{ETS Montreal, University of Quebec}
  \city{Montreal}
  \state{QC}
  \country{Canada}
}
\email{ali.ouni@etsmtl.ca}


\begin{abstract}
The microservices architectural style offers many advantages such as scalability, reusability and ease of maintainability. As such microservices has become a common architectural choice when developing new applications. Hence, to benefit from these advantages, monolithic applications need to be redesigned in order to migrate to a microservice based architecture. Due to the inherent complexity and high costs related to this process, it is crucial to automate this task. In this paper, we propose a method that can identify potential microservices from a given monolithic application. Our method takes as input the source code of the source application in order to measure the similarities and dependencies between all of the classes in the system using their interactions and the domain terminology employed within the code. These similarity values are then used with a variant of a density-based clustering algorithm to generate a hierarchical structure of the recommended microservices while identifying potential outlier classes. We provide an empirical evaluation of our approach through different experimental settings including a comparison with existing human-designed microservices and a comparison with 5 baselines. The results show that our method succeeds in generating microservices that are overall more cohesive and that have fewer interactions in-between them with up to 0.9 of precision score when compared to  human-designed microservices. 
\end{abstract}




\maketitle

\section{Introduction}

The microservices architecture is a set of finely grained and lightweight components that each define a specific business need and that interact with each other through a RESTful API \cite{almarimi2019web,saied2020towards,saied2016cooperative}. It offers multiple advantages such as scalability, independent development, reusability, maintainability and compatibility with cloud technologies \cite{vayghan2021kubernetes, vayghan2018deploying}. Due to these benefits, many software companies have been migrating their legacy monolithic applications into microservices based software such Netflix, eBay, Amazon, IBM, etc. However, this process has proven to be difficult and costly \cite{Henry2020}. 

As the demand for scalable, available and interoperable software has increased, the monolithic architectural style was not enough to answer the needs of modern applications \cite{vayghan2021kubernetes,shatnawi2018identifying, benomar2015detection, vayghan2019microservice, vayghan2019kubernetes}. Hence, the microservices architectural design has become a commonly preferred solution when developing new software applications  \cite{msextractor} . 

One of the major challenges encountered in this endeavor is identifying how to split the components of the monolithic application. These components are more often than not highly cohesive and tightly coupled due to the nature of this architectural design. Another challenge lies within the subjectivity of the process which relies heavily on the opinions of the software experts handling this task. For applications with older frameworks or programming languages, developers that are familiar with the code base are hard to retain due to turnover which adds further to the difficulty of this task \cite{ferreira2020turnover,almarimi2020detection,saied2018towards,saied2015visualization, huppe2017mining}.

For these reasons, prior research has attempted to develop automated tools for extracting a microservices architecture from a monolithic application \cite{bunch,mem,cogcn,code2vecDecomp,msextractor}. Specifically, the main focus of the problem is identifying the different components  that can constitute separate microservices. As such, this problem is handled as a clustering problem where the objective is extracting different clusters (i.e., microservices) from the sample elements (the application’s components) . The existing approaches can be split into two categories based on whether they use static \cite{bunch,msextractor,code2vecDecomp} or dynamic \cite{mono2micro,cogcn,fosci} analysis of the legacy applications. Most of these solutions utilize clustering-based algorithms or evolutionary algorithms. However, these tools often require the number of candidate microservices as input which is on its own a difficult task that requires the calibration of expert developers.

In this paper, we propose a novel microservices extraction approach that utilizes a Hierarchical DBSCAN algorithm \cite{epsilon_dbscan,saied2015mining} based on the static analysis of the legacy applications. More specifically, we combine structural and semantic similarity analysis between classes from the source code and then apply the Hierarchical DBSCAN algorithm in order to obtain a set of candidate microservices. The main advantage of using a density-based clustering algorithm is its ability to identify outliers which represent in this case classes that should be removed or refactored in the microservices architecture. Moreover, the Hierarchical DBSCAN not only provides the extracted microservices, but generates as well a hierarchical view of these microservices which would facilitate their customization and understanding their functional role.

In order to evaluate our approach, we compared the extracted microservices with microservices designed by human developers and reviewed how similar they are for 3 projects.  The median precision score for each project surpassed 0.65 and had a maximum value above 0.9 in 2 of them. We compared as well the quality of the extracted microservices with those generated by other state-of-the-art approaches using 4 different metrics and 4 projects of varying complexity. Our solution achieved the best overall Structural Modularity scores for 3 projects and the best Inter-Call Percentage and Interface Number scores for all projects.

The main contributions of this paper are as follows:
\begin{enumerate}
    \item The formulation of microservices extraction as hierarchical clustering combining structural and semantic analysis. 
     Our solution provides a hierarchical view of the potential microservices and detects possible outliers. 
    \item Introducing novel metrics that enable the comparison of different microservices decompositions.
    \item An Empirical evaluation on our approach's ability to extract  microservices and a comparison with state-of-the-art solutions.
\end{enumerate}

The paper is organized as follows. We present the work related to this research in section 2. Then, we present our proposed solution in section 3. In section 4, we showcase the empirical evaluation of our approach. Afterwards, we discuss the results this work in section 5. Finally, we conclude the paper in section 6 and we outline our future work.

\section{Related Work}
Most solutions in the literature that tackle problems similar to microservices extraction can be split into two distinct steps. 

The first step revolves around the type of input that is fed to the solution and how it is processed. For example, the method proposed by MSExtractor\cite{msextractor}, bunch\cite{bunch} and \cite{code2vecDecomp} take in as input the source code of the monolithic application on which they apply different static analysis techniques. More specifically, both MSExtractor and bunch build call graphs that encode the interactions between the classes in these systems. The method mentioned in \cite{code2vecDecomp} converts the source code into a set of Abstract Syntax Trees which are fed to a code embeddings model \cite{code2vec}. The intuition behind the use of static analysis and more specifically the source code structure is that structurally similar classes or functions should be grouped together. 

Other solutions such as Mono2Micro\cite{mono2micro}, FoSCI\cite{fosci} and CO-GCN\cite{cogcn} are based on the analysis of the use cases and execution traces of the monolithic systems. These solutions aim to group together classes or functions that interact together at runtime for each business need provided with the input. Mono2Micro measures similarity metrics between the classes based on the execution traces. CO-GCN utilizes the execution traces to build matrices as features for its inference architecture. These methods of analysis are not mutually exclusive as they can be combined for better results. For example, MSExtractor analyzes the domain terms included in the classes in addition to the structural analysis with the assumption that each microservices should contain classes with similar domain concepts. Another example is CO-GCN which, in addition to using execution traces, uses the source code of the input application in order to build its model’s architecture. 

On the other hand, there are solutions that use different inputs such as MEM\cite{mem} which analyzes the git commit history of the monolithic applications. This solution constructs a graph from the git history that encodes the similarity between the classes.

The second step of each solution takes in as input the data processed in the previous step and applies on it an algorithm in order to generate the decompositions. Most solutions use clustering algorithms like in \cite{code2vecDecomp} which uses the vectors obtained from the code embeddings as input to an Affinity propagation \cite{affinity} clustering algorithm. Mono2Micro\cite{mono2micro} uses the similarity metrics it measured with an agglomerative single-linkage clustering algorithm \cite{hierarchical}. MEM\cite{mem} introduces its own clustering algorithm based on the graph it generated. 

Some solutions propose search algorithms in order to achieve their task. MSExtractor\cite{msextractor} uses the non-dominated sorting genetic algorithm (NSGA-II) \cite{nsga2} while FoSCI\cite{fosci} uses both NSGA-II and hierarchical clustering on the execution traces. Bunch\cite{bunch} on the other hand uses a different approach which applies a hill-climbing algorithm.

\section{Proposed Approach}
In this section, we detail our proposed approach for extracting candidate microservices from a given monolithic application.

\subsection{Problem Formulation}
The objective of our approach is to identify a set of microservices when a given a legacy monolithic  application which is characterized by the set of its classes. We assume that each class can only be contained within one microservice. As such, we formulate our problem as a clustering task where the microservices correspond to clusters and the classes define the samples.

The input of the solution is the legacy application which is defined as a set of classes $C = \{c_1,c_2,c_3,…,c_N\}$ where N is the number of classes in the system. The output of the solution is $M = \{m_1,m_2,m_3,…m_K\}$ where K is the number of the microservices selected by the solution. 
The output can be represented as a vector $X = [x_1,x_2,x_3,…x_N]$ of size N and where $x_i$ = j means that class $c_i$ is contained in microservice $m_j$.

We also define outlier classes which are too dissimilar from other classes and hence, do not belong to any specific cluster. consequently, we define $x_i = -1$ which means that class $c_i$ is an outlier.

\begin{figure*}[ht]
  \centering
  \includegraphics[width=\textwidth]{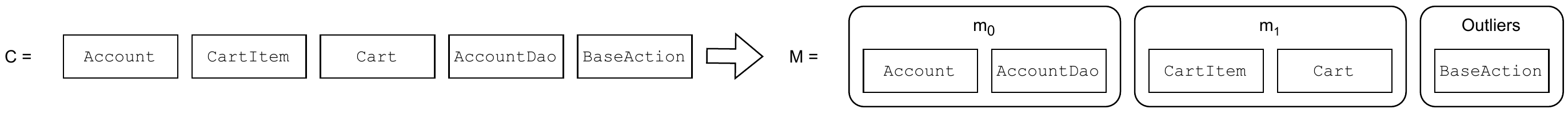}
  \caption{An example showcasing a potential decomposition for an application of 5 classes.}
  \Description{An example showcasing a potential decomposition for an application of 5 classes.}
  \label{jpetstore_example}
\end{figure*}

Figure \ref{jpetstore_example} showcases a simple example of a monolithic application that consists of 5 classes. The second part of the Figure represents a potential decomposition which contains two microservices and an outlier class. In this case, the vector X is represented as $X = [0,1,1,0,-1]$.

In order to apply the clustering algorithms, we need to define a distance function between each class $c_i$ and $c_j$. In the next subsection, we detail this distance function.

\subsection{Representation of the Monolithic Application}

As mentioned in the previous section, most of the state-of-the-art solutions \cite{bunch,msextractor,code2vecDecomp} rely on the static analysis of the source code of the monolithic application in order to extract meaningful patterns. In our approach, we utilize a static analysis of the source code in order to extract the relations and dependencies between the different classes. We define two distinct types of similarities between classes as follows: 

\textbf{Structural Similarity ($Sim_{str}$)}: This metric is based on the shared number of method calls between 2 classes. It encodes the dependency between them and as such evaluates the similarity from a functional point of view  \cite{msextractor,saied2015could}. The objective of grouping together classes based on this similarity is to obtain microservices that are cohesive from the implementation perspective. 
For two given classes $c_i$ and $c_j$, the structural similarity is defined as follows:

\begin{equation}
\resizebox{.9\hsize}{!}{

    $Sim_{str}(c_i,c_j)=\left\{
    \begin{matrix}
    \frac{1}{2}\left ( \frac{calls(c_i,c_j)}{calls_{in}(c_j)}+\frac{calls(c_j,c_i)}{calls_{in}(c_i)} \right )
 & if \; calls_{in}(c_i)\neq0 \; and \;  calls_{in}(c_j)\neq 0\\ \frac{calls(c_i,c_j)}{calls_{in}(c_j)}
 & if \; calls_{in}(c_i)=0 \; and \;  calls_{in}(c_j)\neq 0\\ \frac{calls(c_j,c_i)}{calls_{in}(c_i)}
 & if \; calls_{in}(c_i)\neq0 \; and \;  calls_{in}(c_j)= 0\\ 0
 & otherwise
 
\end{matrix}
\right. $
}
\end{equation}

\noindent where $calls(c_i,c_j)$ represents the number of times a method of the class $c_i$ has called a method from the class $c_j$. On the other hand, $call_{in}(c_i)$ is the number of calls incoming to $c_i$.

The values of $Sim_{str}(c_i,c_j)$ are in the range [0,1] where 1 indicates classes $c_i$ and $c_j$ are very similar and used exclusively together and 0 indicates that they are completely independent.

\textbf{Semantic Similarity ($Sim_{sem}$)}: This metric utilizes natural language processing in order to measure how related are the domain semantics of the two given classes \cite{msextractor,saied2015could}. 
Given that a microservice should provide a specific function and/or a single domain use case, we need to identify classes that serve similar domain use cases. Assuming the monolithic projects were coded using standard business practices where class, method and variable names reflect business concepts and comments describe their function, analyzing the terminology used within these components can be a powerful tool for extracting the domain significance of each class.

Hence, each class is defined by the set words in its comments, parameter names, field names, method name and variable names. Each word is preprocessed includig splitting using CamelCase, filtering out stop words, and stemming. In the CamelCase splitting phase, we take as input the method, variable and member names and split them into different words assuming that the CamelCase naming convention is used. For example, the word CamelCase would become a list: [camel, case]. Next, we filter out stopwords. Finally, we apply stemming which removes parts of the words in order to remove the impact of conjugation. For example, the words 'stemming' and 'stemmed' would become 'stemm'.  The final result is a vector of size $n_V$ where $n_V$ is the number of words in the domain vocabulary extracted from the monolithic application. This vector is measured using a TF-IDF (Term Frequency-Inverse Document Frequency) model \cite{tfidf}. 
The Semantic Similarity metric is defined as the cosine similarity between two classes:

\begin{equation}
    Sim_{sem}(c_i,c_j)= \frac{\vec{c_i}\cdot\vec{c_j} }{\left \| \vec{c_i} \right \|\left \| \vec{c_j} \right \|}
\end{equation}

\noindent where $\vec{c_i}$ and $\vec{c_j}$ are the TF-IDF vectors of class $c_i$ and class $c_j$ and $\left \| \vec{c_i} \right \|$ is the Euclidian norm of vector $\vec{c_i}$.

The values of $Sim_{sem}$ range between 0 and 1 where the value 1 signifies that both classes use the exact same vocabulary.

\textbf{Class Similarity ($CS$)}: The previous similarity metrics represent different aspects of the relations between classes. These two metrics do not necessarily correlate and as such using only a single one of them does not guarantee satisfying the other. For these reasons, we use the Class Similarity metric \cite{msextractor} which represents a weighted sum between the Structural similarity and Semantic similarity of two given classes:
\begin{equation}\label{eq_CS}
    CS(c_i,c_j)= \alpha \, Sim_{str}(c_i,c_j)+ \beta \, Sim_{sem}(c_i,c_j)
\end{equation}

\noindent where $\alpha \in [0,1]$, $\alpha + \beta = 1$ and $CS \in [0,1]$

\subsection{Clustering Algorithms}
Given the Class Similarity metric, we can apply a clustering algorithm to extract the microservices. More specifically, we utilize the DBSCAN algorithm \cite{dbscan}. 

DBSCAN is a density based clustering algorithm where the aim is to group together points that are densely packed in the search space and identify noisy points which do not fit into any clusters using two main concepts which are the neighborhood distance and the minimum number of points in a neighborhood  \cite{dbscan}. 

The advantage of using DBSCAN in this case is that defining the number of target microservices is no longer a requirement since the algorithm identifies on its own the existing clusters. 
The second advantage is the minimum number of sample hyper-parameter which enables the addition of minimum number of classes per microservice constraint. This constraint helps with preventing the extraction of extremely small microservices which usually do not have a practical significance. However, this parameter does not help with dealing with extremely large microservices.
The final advantage of applying this algorithm is the detection of noisy points which we consider as outlier classes. They are classes that are not similar enough to a minimum number of classes in original application and as such cannot be included in any of the extracted microservices.

However, DBSCAN is still constrained by another hyper-parameter which is the neighborhood distance $\epsilon$. The resulted number of extracted microservices, the number of outlier classes and the size of clusters are heavily dependent on this hyper-parameter. For this reason, we used in our solution the variant of the DBSCAN algorithm defined in \cite{epsilon_dbscan} which was called $\epsilon$-DBSCAN. 

The first step in this algorithm is to cluster the input data using DBSCAN with an $\epsilon$ value equal to 0. This generates the initial cluster which contains identical points for a given metric. Afterwards we increment $\epsilon$ with a certain value that is defined within the input. We apply DBSCAN once more but with the new $\epsilon$ value which generates a new layer of clusters that contain within them the previous clusters as well as potentially new points. We continue this step until $\epsilon$ is equal to MaxEpsilon which is defined by the user.

Although $\epsilon$-DBSCAN has the MapEpsilon hyper-parameter which influences the final layer of the extracted microservices, the main advantage of this algorithm is the hierarchical nature of its results. The output of $\epsilon$-DBSCAN is a set of ordered layers that correspond each to the microservices extraction at an epsilon step. This output can be used to visualize the evolution of microservices and their classes at each step and allows for the differentiation between very similar classes and those at the edges of the microservices. For a domain expert, this can be a powerful tool to discover outlier classes as well and the optimal point to define the extracted microservices.

\begin{figure*}[ht]
  \centering
  \includegraphics[width=0.8\textwidth]{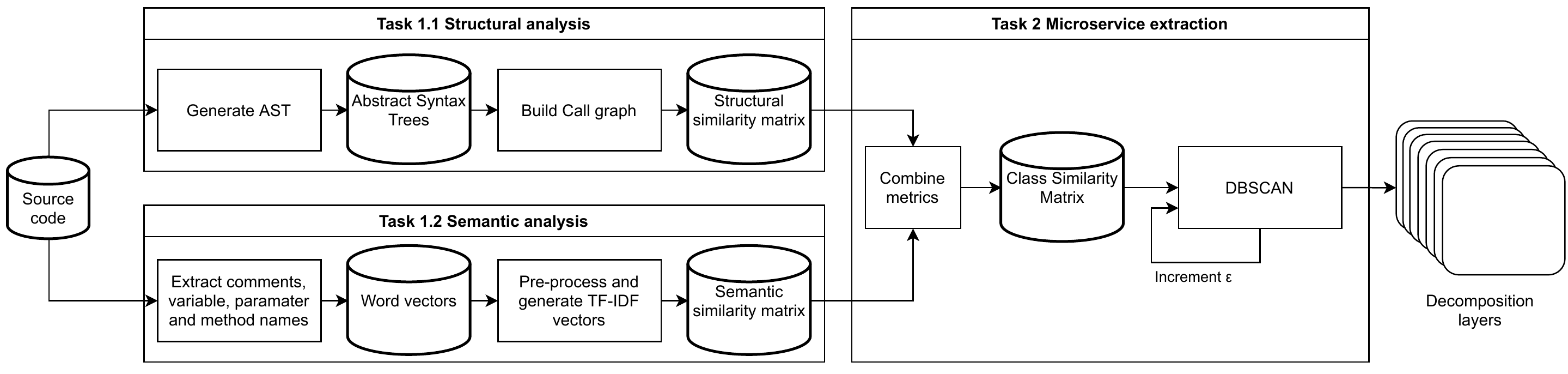}
  \caption{A summary of the steps taken to extract the microservices.}
  \Description{A summary of the steps taken to extract the microservices.}
  \label{method_process}
\end{figure*}

Figure \ref{method_process} summarizes the process executed in order to extract the microservices using our approach. Through this Figure, we can see that as we take the source code of input, two tasks start in parallel. In the first task, we use the Abstract Syntax Trees to build the call graph that encodes the interactions between the classes. Afterwards, we generate the Structural similarity matrix. The second task extracts the comments, method, parameter and variable names within the classes, applies the Natural Language pre-processing like CamelCase splitting, stop-word removal and stemming and generates the TF-IDF vectors. These vectors are used to measure the semantic similarity matrix. By combining the output of both of these tasks, we generate the class similarity matrix and start the extraction task with $\epsilon$-DBSCAN. The output of this phase is a set of decompositions that reflect the output of DBSCAN at each epsilon step. The final layer represents the extracted microservices while the rest can be used for in-depth analysis and manual customization of the microservices. 

\section{Evaluation}

In this section, we evaluate the effectiveness of our approach in extracting the proper microservices.  We first describe the research questions we are seeking to investigate. Thereafter, we showcase and discuss the experimental results. 

\subsection{Research Questions}
We designed our experimental study to answer the following research questions (RQs):
\begin{itemize}

    \item \textbf{Q1:} How well do the extracted microservices compare to those that were manually identified by software engineers. 
    \item \textbf{Q2:} What is the impact of various experimental settings on the extracted microservices quality?
    \item \textbf{Q3:} How does our solution perform when compared with state-of-the-art microservices extraction baselines?
\end{itemize}

\subsection{Evaluation and results for RQ1}
\subsubsection{Evaluation protocol}

To answer RQ1, we selected 3 Open-Source microservices-based Java projects with different degrees of complexity, namely \textit{Spring PetClinic}\footnote{https://github.com/spring-petclinic/spring-petclinic-microservices}, \textit{Microservices Event Sourcing}\footnote{https://github.com/chaokunyang/microservices-event-sourcing} and \textit{Kanban Board demo}\footnote{https://github.com/eventuate-examples/es-kanban-board}. The details of these projects can be viewed in the Table \ref{micro_projects}.

\begin{table}[]
\centering
\caption{Metadata of the Microservice-based projects.}
\resizebox{\columnwidth}{!}{\begin{tabular}{@{}lllll@{}}
\toprule
Project                      & Version & SLOC & \# of classes & \# of microservices \\ \midrule
Spring PetClinic             & 2.3.6   & 1,889 & 43            & 7                   \\
Microservices Event Sourcing  & 2.8.0   & 4,597 & 121           & 12                  \\
Kanban Board                 & 0.1.0   & 4,380 & 118           & 21                  \\ \bottomrule
\end{tabular}}

\label{micro_projects}
\end{table}

To evaluate our approach, we need to define measures that can compare two given sets of classes that represent the set of extracted microservices and the ground truth microservices. It is worth noting that the encoding of the clusters for any given decomposition solution can be different. Let's take the application shown in Figure \ref{jpetstore_example} as an example. Let's suppose that the original set of microservices was $X_t = [0,0,1,1,2]$. If we get a decomposition denoted by  $X_1 = [1,1,2,2,0]$, we will have two vectors that are different but encode the same decomposition. This issue becomes more complicated if the vectors have different microservices (for example $X_2 = [0,1,1,1,2]$). In such a case, it is even more ambiguous which microservices correspond to each other. 

To overcome this challenge, we first need to identify the corresponding ground truth microservice for each extracted microservice. As such, we define the corresponding microservice as the microservice that has the largest number of common classes with the extracted microservice. We introduce the function \ref{corr} that given an extracted microservice $m_i$ and a different set of microservices M (set of ground truth microservices), $Corr(m_i,M)$ selects the ground truth microservice with the largest number of common classes with the extracted microservice.

\begin{equation}
\label{corr}
    Corr(m_i,M) = \underset{m_j\in M}{argmax}(\frac{\left | m_i \cap m_j \right |}{\left | m_i \right |})
\end{equation}

We then can calculate the precision metric, defined in equation \ref{precision}, which is the mean of the percentage of the correctly identified classes out of the total number of identified classes for each extracted microservice.
\begin{equation}
\label{precision}
    precision = \frac{1}{\left | M \right |}\times \underset{\forall m_i\in M}{\sum } \frac{\left | m_i \cap Corr(m_i, M_t) \right |}{\left | m_i \right |}
\end{equation}
where M is the set of the extracted microservices, $M_t$ is the set of the original or ground truth microservices and $Corr(m_i,M_t)$ is the microservice from $M_t$ that corresponds to the extracted $m_i$ .

We also calculate as well the Success Rate (SR) that measures the percentage of successfully retrieved microservices based on the precision metric:
\begin{equation}
\label{SR}
    SR = \frac{1}{\left | M \right |}\times \underset{\forall m_i\in M}{\sum } matching(m_i, Corr(m_i, M_t))
\end{equation}
\noindent where $matching$ is defined as:
\begin{equation}
\label{matching_value}
    matching(m_1,m_2) = \begin{cases}
                    1 & \text{ if } \frac{\left | m_1 \cap m_2 \right |}{\left | m_1 \right |} \geqslant threshold\\ 
                    0 & \text{ otherwise } 
                \end{cases}
\end{equation}
\noindent where $m_1$ and $m_2$ are two sets of classes and $threshold \in \left [ 0,1 \right ]$. So for a given $threshold = \frac{k}{10}$ we calculate the $SR@k$ of the corresponding k value.

For each of these test projects, we regrouped all of their Java classes together in order to simulate a Monolithic architecture. This version served as input for our solution while its original version was considered as the true decomposition with which we compare our results. For each of the mentioned projects, we generated multiple microservices decompositions with varying hyper-parameter values. Afterwards we calculate the metrics precision, SR@5, SR@7 and SR@9 based on these decompositions and the corresponding true decomposition.

\begin{figure*}[ht]
  \centering
  \includegraphics[width=\textwidth]{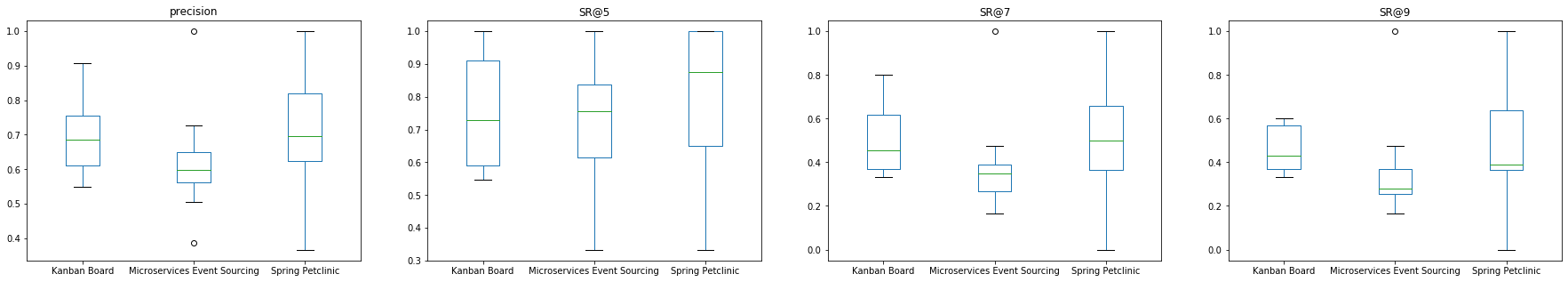}
  \caption{Boxplots showcasing the comparison between the generated decompositions and their corresponding true microservices.}
  \Description{Boxplots showcasing the comparison between the generated decompositions and their corresponding true microservices. \aymen{export figures as pdf instead of png}}
  \label{boxplot_micro}
\end{figure*}

\subsubsection{Results}

The Figure \ref{boxplot_micro} showcases the results in the form of boxplots for each project and each metric. We can observe that the median precision for each project is within the range [0.6,0.7]. More specifically, both Kanban Board demo and Microservices Event Sourcing achieved precision values higher than 0.5 for all decompositions except for a single outlier. On the other hand, there’s a large variability in the precision values obtained for Spring Petclinic. This difference shows that as the number of classes increases, the stability of the results improves. 
As for the success rate, we can observe that as we increase the threshold, the median and minimum scores achieved for each project drop but we can as well see that for Kanban Board demo and Microservices Event Sourcing, the values have less variance. Additionally, there is not a significant difference between the scores for SR@7 and SR@9. These results suggest that for most decompositions, a high percentage of their microservices achieve a precision score higher than 0.5 and that a significant number among them have a high precision score that exceeds 0.9. We can see as well the same pattern of the variance of the results decreasing when the size of the project increases.

The microservices extracted for the project Microservices Event Sourcing seem to have overall lower scores than the other projects. The most likely explanation, however, stems from the fact that this projects does not use a single natural language for its domain terms unlike the other projects which utilize only the English language. This serves as an additional roadblock for this project. Nonetheless, the results achieved are comparable to the others and the precision values are in the same range as the others.

\begin{figure*}[ht]
  \centering
  \includegraphics[width=\textwidth]{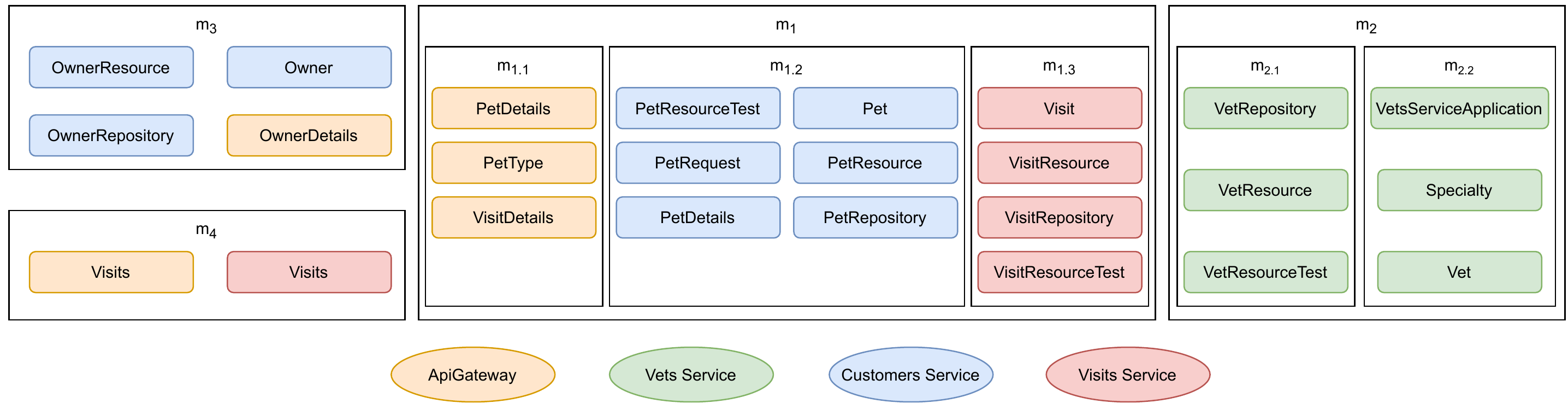}
  \caption{A subset of the microservices obtained from a decomposition of the project Spring PetClinic.}
  \Description{A subset of the microservices obtained from a decomposition of the project Spring PetClinic.}
  \label{rq1_example}
\end{figure*}

As an example, we focus on a decomposition of the project Spring PetClinic which is an online platform that provides veterinary services that was implemented with a microservices architecture in mind. Given the size of the project, the Figure \ref{rq1_example} shows a subset of the microservices obtained from one of the decompositions. In this Figure, the ellipses represent the names of some of the original microservices while the large white rectangles represent the new microservices. Within them, we find the classes which are colored based on their original microservices. 

The original microservice mostly implement domain specific concepts like customers, visits and Veterinaries. Other microservices like \texttt{ApiGateway}, \texttt{DiscoveryServer}, \texttt{ConfigServer} and \texttt{AdminServer} implement technical requirements for the used platform. More specifically, the \texttt{ApiGateway} microservice includes a large range of utility classes as well as data classes that have to mirror their counterparts in the other microservices. For this reason, we can observe some classes like Visits which are included twice.

We can see that the microservice $m_2$ contains 6 classes that were originally in the Vets service and that represent utility classes related to the \texttt{Veterinary} domain concept which shows an example of semantically similar classes being grouped together. A second example for such a case is the microservice $m_3$ which contains 4 classes, 3 of which were originally within the Customers microservice and that represent the Owner concept. We can see as well that the 4th class in this case represents as well the Owner concept despite being from a different microservice originally. The microservice $m_4$ shows an example of two classes from different microservices but that mirror the same class being grouped together. In this case, they are the \texttt{Visits} domain object class from the visits microservice and the\texttt{ Visits DTO} from \texttt{ApiGateway}. Finally, we can observe that the largest microservice $m_1$ is grouping multiple classes that represent two major domain concepts: \texttt{Pet} and \texttt{Visit}. Most of the classes in relation to the visits concept in the \texttt{Visits} microservice and the \texttt{ApiGateway} microservice were included in $m_1$. Similarly, most of the concepts in relation to the \texttt{Pet} service in the \texttt{customers} microservice and the \texttt{ApiGateway} microservice were included as well. 

Through this example, we can observe the hierarchical structure of the microservices. the microservice $m_1$ contains in fact 3 potential microservices which are $m_{1.1}$, $m_{1.2}$ and $m_{1.3}$. In this example, our proposed approach extracted initially these 3 microservices with highly similar classes. Since the similarity values in-between these microservices are within the thresholds defined in the input, they were grouped together in a larger microservice that satisfies the conditions defined at the start of the process. We can observe another hierarchy within the microservice $m_2$. In this case, $m_{2.1}$ and $m_{2.2}$ both contain classes that were originally from the Vets service.

\smallskip
{\centering
\fbox{
\begin{minipage}{\linewidth}

The results obtained show that the extracted microservices achieve a median precision score around 0.65 with a maximum value surpassing 0.9 for some projects. Although the extracted microservices are not identical to the original microservices, we observed that they include nonetheless most of the classes from the human-built microservices. Additionally, the hierarchical structure of the results helps the developer with identifying alternative decompositions.
\end{minipage}}}

\subsection{Evaluation and results for RQ2}
\subsubsection{Evaluation protocol}

In the previous question, we observed that for certain projects, the used hyper-parameters can have a large impact on the quality of the result. For example, for Spring Petclinic, there was a large variance in the evaluation metrics. This research question focuses on analyzing the impact of the hyper-parameters of our proposed method on the quality of the final result.  
In order to address this question, we experiment with varying the values of the hyper-parameters MaxEpsilon, $\alpha$ and MinSamples.
\begin{itemize}
    \item \textbf{MaxEpsilon} refers to the maximum value of the $\epsilon$ hyper-parameter that is fed to the DBSCAN algorithm during the $\epsilon$-DBSCAN algorithm. Its values range from 0 to 1 in this case which correspond to the minimum and maximum possible distance between two classes for the metric CS defined in equation \ref{eq_CS}. 
    \item \textbf{alpha $\alpha$}, on the other hand, refers to the weight associated with the similarity value $Sim_{str}$ when calculating CS in the formula \ref{eq_CS}. $\alpha$ values range from 0 to 1 where 0 indicates complete reliance on the $Sim_{str}$ metric which means the extraction is exclusively based on the structural similarity of the classes. On the other hand, $\alpha=1$ corresponds to an extraction process based on the semantic domain similarity between the classes so $CS = Sim_{sem}$.
    \item \textbf{MinSamples} is the minimum number of possible classes that can exist within a microservice. It is a hyper-parameter of the DBSCAN algorithm and in consequence the $\epsilon$-DBSCAN algorithm.
\end{itemize}

For each hyper-parameter, we used the following process: First, we define the range of possible values and the iteration step. Then, we fix the rest of the hyper-parameters. Afterwards, we run our approach for each possible value and we record the extracted microservices. Next, we evaluate the obtained results using different metrics. Finally, we plot the metric values at each step. The following analysis is focused on the Monolithic project JPetStore as it serves as a standard benchmark for this problem and an ideal example to study. Details regarding this project can be viewed in the Table \ref{mono_projects}.

In order to analyze different aspects of the obtained decompositions without the need for the ground truth microservices, we use the following metric:
\begin{itemize}
    \item \textbf{Structural Modularity (SM)}: \cite{fosci} is a metric that combines the measures for the cohesion in each microservice and the coupling between different microservices in order to evaluate the structural quality of the obtained microservices. It is defined as follows: 
    \begin{equation}
        SM = \frac{1}{K}\sum_{i=1}^{K}\frac{\mu_i}{m_i^{2}} \: - \: \frac{1}{(K(K-1))/2}\sum_{i\ne j}^{K}\frac{\sigma_{i,j}}{2\, m_i\, m_j}
    \end{equation}
    Where K is the number of the extracted microservices, $\mu_i$ is the number of unique calls between the classes in microservice i, $m_i$ is number of classes in microservice i and $\sigma_{i,j}$ is the number of unique calls between classes of microservice i and classes of microservice j. 
    
    Higher values of SM reflect higher cohesiveness and lower coupling so better structural quality.

    \item \textbf{Interface Number (IFN)}:\cite{msextractor} is a metric for measuring the number of interfaces within a list of extracted microservices. It is defined as:
    \begin{equation}
        IFN = \frac{1}{K}\sum_{i=1}^{K}\left | I_i \right |
    \end{equation}
    Where K is the number of the extracted microservices, $I_i$ is the set of interface classes within microservice i. An interface class in this case is a class that has been called by a class in an external microservice. 
    
    Lower values of IFN indicate a better result.

    \item \textbf{Non-Extreme Distribution (NED)}:\cite{mono2micro} One of the most notable problems encountered in microservice extraction is the Boulder and Grain problem where solutions tend to output microservices that are extremely large (Boulders) or extremely small (Grain). This metric enables the detection and evaluation of such cases. It is defined as: 
    \begin{equation}
        NED = 1-\frac{\left |\{m_i\; ; \: 5<\left | m_i \right |<20,i \in [1,K]\}  \right |}{K}
    \end{equation}
    Where K is the number of the extracted microservices and $|m_i|$ is the size of microservice $m_i$. 
    
    Lower values of NED indicate lower number of extreme microservices which corresponds to better results. 

    \item \textbf{Inter Call Percentage (ICP)}:\cite{mono2micro} is measured in the literature as the percentage of runtime calls between two microservices. In this case, we are measuring ICP as the percentage of static calls between two microservices. The objective of this metric is still the same which is evaluating the dependencies between the microservices or in other terms the coupling. It is defined as follows:
    \begin{equation}
        ICP = \frac{\sum_{i=1,j=1,i\ne j}^{K}icp(M_i,M_j)}{\sum_{i=1,j=1}^{K}icp(M_i,M_j)}
    \end{equation}
    Where K is the number of microservices, $M_i$ is the set of classes in microservice i and $icp(M_i,M_j)$ is defined as: 
    \begin{equation}
        icp(M_i,M_j) = \sum_{c_k \in M_i}\sum_{c_l \in M_j}\left (log(calls(c_k,c_l))+1  \right )
    \end{equation}
    where $calls(c_k,c_l)$ is the number of calls from class $c_k$ to class $c_l$.
    
    Lower numbers of ICP indicate less interactions between the different microservices which represents a better result.
\end{itemize}

\subsubsection{Results}

\begin{figure*}[ht]
  \centering
  \includegraphics[width=0.7\textwidth]{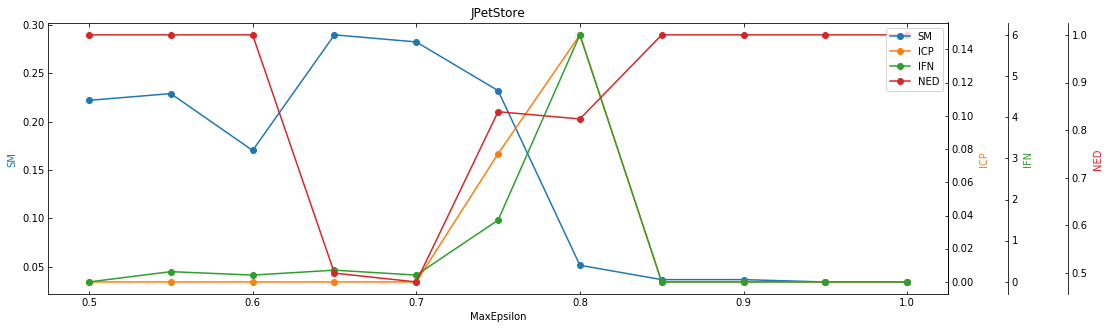}
  \caption{Evaluation metrics for different MaxEpsilon values when extracting microservices from the project JPetStore}
  \Description{Evaluation metrics for different MaxEpsilon values when extracting microservices from the project JPetStore}
  \label{fig_maxepsilon}
\end{figure*}
The following subsection showcases the results when individually varying each hyper-parameter:

\textbf{MaxEpsilon}: We varied the MaxEpsilon values from 0 to 1 with a step equal to 0.05 . We fixed alpha to 0.5 and MinSamples to 2. The Figure \ref{fig_maxepsilon} shows the results obtained for the project JPetStore. We can observe that for MaxEpsilon values below 0.5, all of the classes have been classified as outliers. On the other hand, we can see that as we increase MaxEpsilon, the SM value, which is plotted in blue, starts off by slightly increasing and then rapidly decreasing starting from the value 0.8 where we can observe a large spike in the values of IFN and ICP. The plots for these couple of metrics, which correspond respectively to the green and orange plots, share the same shape where they start out with low values, rapidly increase between 0.7 and 0.8 and finally decrease back to almost 0. As for NED, which is represented in red, starts with high values for the range of MaxEpsilon in [0.5, 0.6]. Then, it drops to 0.5 and starts increasing slowly back to 1 when MaxEpsilon ranges between 0.65 and 0.85. Finally, the NED values stay constant for the rest of the range of MaxEpsilon values.

Observing these metrics together, we can infer the explanation behind these plot shapes and select the ideal MaxEpsilon value for this project. MaxEpsilon is the hyper-parameter that controls how similar are the classes that we are grouping together to form the microservices. In other terms, ranging its values from 0 to 1 is akin to loosening the condition for grouping the classes.

So, not having any microservices for MaxEpsilon values between 0 and 0.5 shows that the condition is too strict in that case and the algorithm is unable to find any classes that satisfy it. As we continue to increase MaxEpsilon, the algorithm starts gradually grouping together some classes that satisfy the corresponding condition which explains the observation in the range [0.5, 0.6] where NED and SM are high due to having small but coherent microservices. For the values 0.65 and 0.7, we can observe higher SM values, lower NED values and low ICP and IFN values due to having larger but more balanced microservice and that still contain coherent classes. As for the values 0.75 and 0.8, the sharp increase in NED, ICP and IFN values and the sharp decrease of SM values suggest that the condition is now loose enough that outlier classes are added to the microservices which results in increased inter-microservices interactions and less coherent microservices. As for the rest of the values, the high NED value and the low values for the rest of the metrics suggest that the condition is too loose and the microservices have been grouped together which is to be expected for extreme MaxEpsilon values.

From this Figure, it is clear that the best MaxEpsilon values for the project JPetStore would be 0.65 or 0.7 where the trade-off between the metrics would be at its best. For the rest of the projects, we can observe a similar pattern as the one we described. 

\begin{figure*}[ht]
  \centering
  \includegraphics[width=0.7\textwidth]{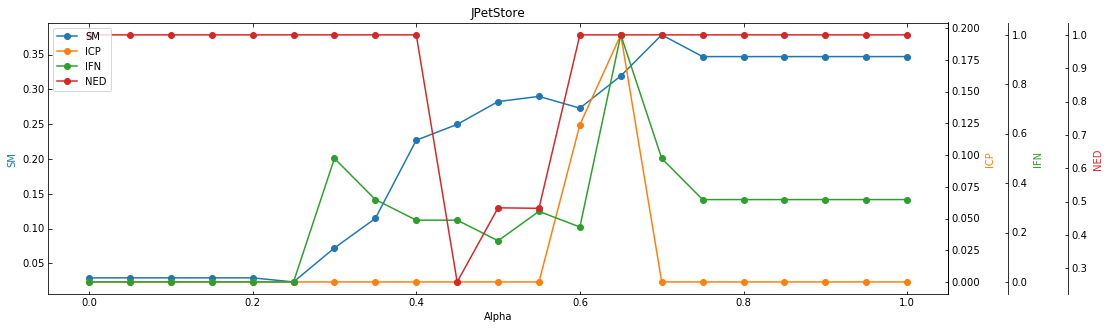}
  \caption{Evaluation metrics for different alpha values when extracting microservices from the project JPetStore}
  \Description{Evaluation metrics for different alpha values when extracting microservices from the project JPetStore}
  \label{fig_alpha}
\end{figure*}

\textbf{alpha $\alpha$}: For this experiment, we fixed MaxEpsilon to 0.7 based in the previous results and MinSamples to 2. We varied alpha from 0 to 1. Figure \ref{fig_alpha} shows the results obtained for the project JPetStore. We can observe that overall, SM and IFN values seem to be increasing when we increase alpha. On the other hand, ICP shows only a couple of spikes for alpha equal to 0.6 and 0.65. As for NED, its values stay equal to 1 for all alpha values except for the range [0.45, 0.55] where it drops to 0.25 and 0.45. 

In this case, varying alpha from 0 to 1 is equivalent to relying less and less on the structural similarity and increasing the impact of the semantic similarity between the classes. This can explain why ICP and IFN values are higher for high alpha values as relying more on the structural similarity will lead to the algorithm prioritizing grouping together classes that are more similar structurally and hence that communicate with each other which in turn lowers these metrics.

From this Figure, we can conclude that the best values for alpha should be in the range [0.45, 0.55] where we balance the impact of both similarity values. This was the case as well for the other projects where this range represented the base trade-off point between the different metrics.

\begin{figure*}[ht]
  \centering
  \includegraphics[width=0.7\textwidth]{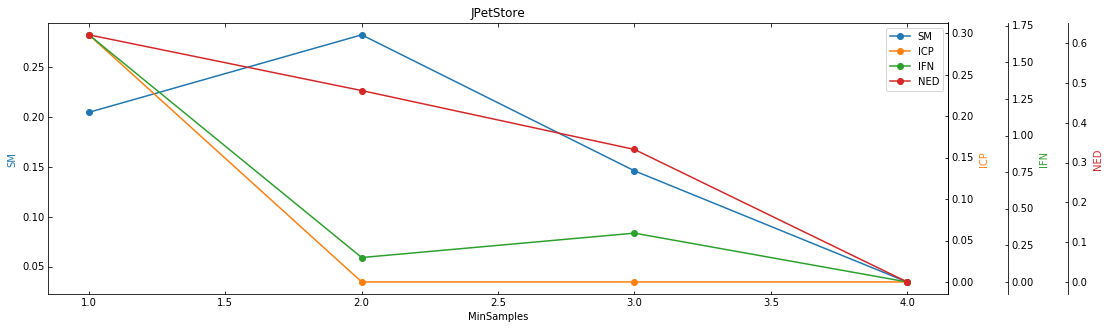}
  \caption{Evaluation metrics for different MinSamples values when extracting microservices from the project JPetStore}
  \Description{Evaluation metrics for different MinSamples values when extracting microservices from the project JPetStore}
  \label{fig_MinSamples}
\end{figure*}

\textbf{MinSamples}: In this case, we fixed MaxEpsilon to 0.7 and alpha to 0.5 and we varied the MinSample hyper-parameter in the range [1, 4]. The Figure \ref{fig_MinSamples} shows the results for the project JPetStore. 

We see that ICP, IFN and NED decrease as we increase MinSample and that SM peaks at the value 2 and decreases afterwards. We observed similar results for the rest of the projects which suggests that the best value for this hyper-parameter is 2.

\smallskip
{\centering\fbox{
\begin{minipage}{\linewidth}

The optimal values for MaxEpsilon are in the range [0.65, 0.75]. For the hyper-parameter $\alpha$, they are in the range [0.45, 0.55]. Finally, for MinSample, the optimal value is 2.
\end{minipage}}}

\subsection{Evaluation and results for RQ3}
\subsubsection{Evaluation protocol}
For this research question, we compared our solution with 5 baselines that tackle similar problems but with different methods which are Bunch\cite{bunch}, CoGCN\cite{cogcn}, FoSCI\cite{fosci}, MEM\cite{mem} and Mono2Micro\cite{mono2micro}. For each of these baselines as well as our solution, we measured the quality of the obtained microservices using the 4 metrics SM, IFN, ICP and NED.

In order to compare these State-of-the-art solutions, 4 Open-source monolithic Java projects were used as subjects for testing and evaluating the performance of our approach which are JPetStore\footnote{https://github.com/KimJongSung/jPetStore}, DayTrader\footnote{https://github.com/WASdev/sample.daytrader7}, Plants\footnote{https://github.com/WASdev/sample.mono-to-ms.pbw-monolith} and AcmeAir\footnote{https://github.com/acmeair/acmeair}. The Table \ref{mono_projects} lists these projects as well as their metadata. 

\begin{table}[]
\centering
\begin{tabular}{@{}llll@{}}
\toprule
Project   & Version & SLOC   & \# of classes \\ \midrule
JPetStore & 1.0     & 3,341  & 73            \\
DayTrader & 1.4     & 18.224 & 118           \\
Plants    & 1.0     & 7,347  & 40            \\
AcmeAir   & 1.2     & 8,899  & 86            \\ \bottomrule
\end{tabular}
\caption{Metadata of the Monolithic projects.}
\label{mono_projects}
\end{table}
However, since each solution has at least one hyper-parameter that can significantly impact the quality of the final result, we extracted different microservices decompositions from each solution based on different values of their corresponding hyper-parameters. For example, Mono2Micro takes as input the number of target microservices, so we generated a different decomposition for each value of this hyper-parameter in [3,5,7,9,11] for the project JPetStore. As for our solution, given the results we observed in the previous section, we decided to fix alpha to 0.5 and MinSample to 2 since these hyper-parameters show less variability between projects when fixed in this range but we varied the values of MaxEpsilon between 0.5 and 0.9 with a 0.05 step. 

\subsubsection{Results}

\begin{figure*}[ht]
  \centering
  \includegraphics[width=0.95\textwidth]{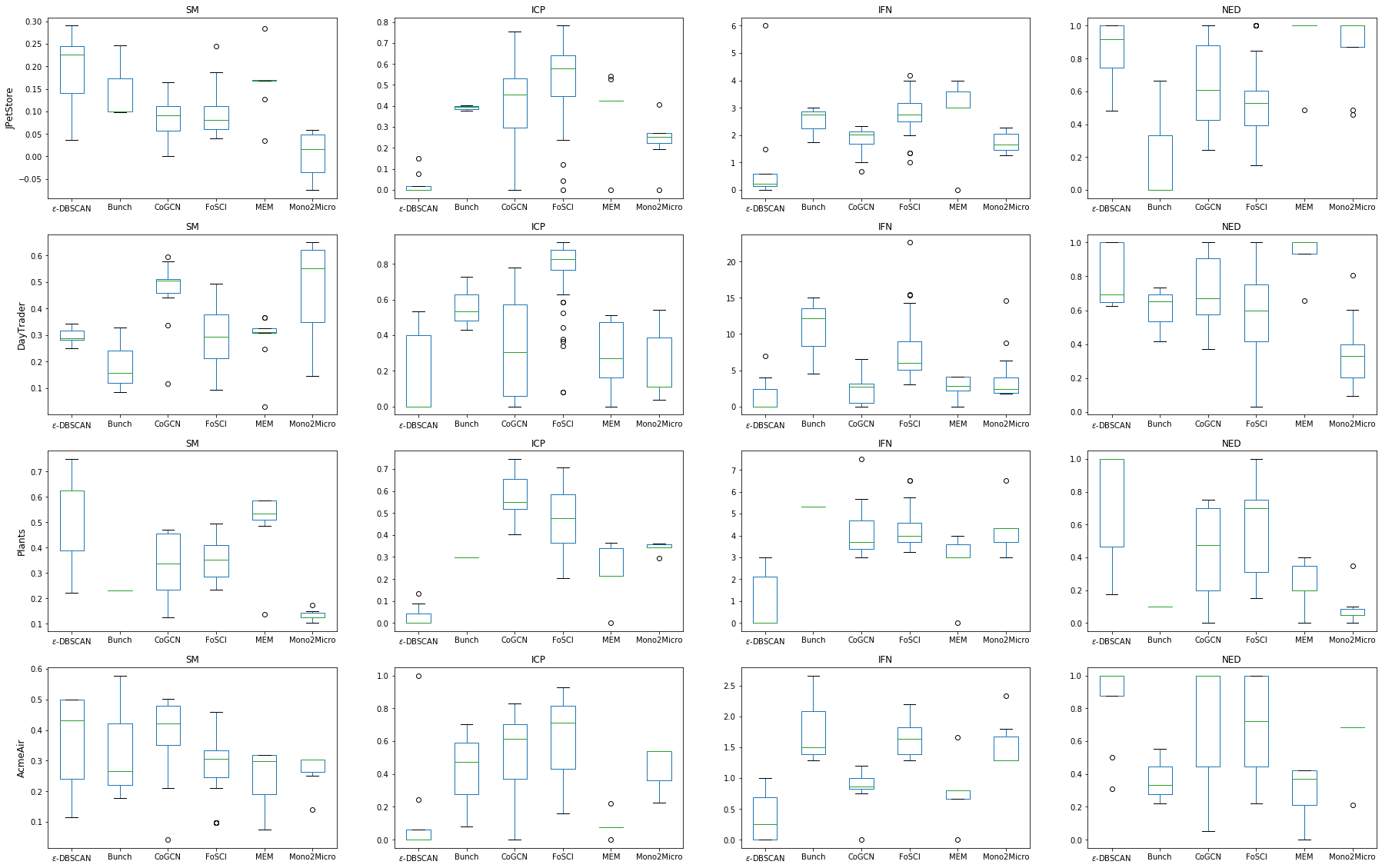}
  \caption{Boxplot representation for the results of each project/baseline/metric combination}
  \Description{Boxplot representation for the results of each project/baseline/metric combination}
  \label{fig_compare}
\end{figure*}

The Figure \ref{fig_compare} shows the boxplot results for each project and metric combination. 

Looking at the metric SM in the first column, we can see that our method outperforms the other baselines for all projects expect DayTrader where we can see that our results always have the highest mean and the highest maximum with one exception being Bunch having a higher maximum for the project AcmeAir. As for DayTrader, our solution outperforms Bunch and has comparable median values to Fosci and MEM.

As for the metrics ICP and IFN which are respectively represented in the second and third columns, we can observe that our solution consistently achieves better results than the other baselines with only one exception where for the project JPetStore we can see one outlier that has higher IFN score than the others. We can see as well that for most cases, the variance of these metrics for our solution is small which suggests that our method often results in decoupled microservices. 

Finally, for the metric NED, our solution seems to outperform MEM and Mono2Micro for the project JPetStore. In the case of DayTrader, it outperforms MEM again and achieves comparable results to most of the rest with the exception of Mono2Micro. For Plant, we can observe a large variance in the results obtained which suggests that choosing the right MaxEpsilon value for this project can have a large impact on the final decomposition. In the previous research question, we recommend the optimal values for this parameter. However, the role of the human experts remains crucial to validate and calibrate the extracted microservices, thus the hierarchical structure of the extracted microservices will be very helpful in this process. Overall, for this metric, we can see that Bunch achieves better NED scores than the rest.

\smallskip
{\centering\fbox{
\begin{minipage}{\linewidth}

The comparison results show that our solution achieves noticeably better results than the  the baseline approaches for the metrics SM, IFN and ICP, but at the cost of higher NED values. 
\end{minipage}}}

\section{Threats to Validity}
The major threat to the validity of our experiments is related the external validity.  Our empirical evaluation was based on a total of 7 open source projects with different architectures and sources. To better generalize the results of our approach, a larger number of projects should have been considered. To mitigate this issue, we selected projects with varying scales and from diverse origins. Future replications of our approach on other monolithic systems are needed. The second threat to validity lies within the qualitative evaluation. 
An alternative solution to evaluate the quality of the extracted microservices would be to involve experienced software engineers who are familiar with microservices-based software systems and the microservices extraction task to inspect the results and give their feedback. We are thus planning to further evaluate our approach with developers in an industrial setting as part of our future work.

As for internal threats to validity, it concerns factors that could influence our observations. The comparison of our solution with the different baselines was based on the results that were generated using specific hyper-parameters which could be an important internal threat to validity. In order to mitigate this issue, we based the results on multiple runs while varying hyper-parameter values for every solution. In a situation without any constraints, a better alternative would be to optimize the hyper-parameters for each solution and experimental setup. The used performance metrics represent a threat to validity. To mitigate this issue, we employed four different metrics that reflect different aspects such as the cohesion within the microservices, the interactions between the microservices and their granularity. Other metrics can also be considered such as the number of microservices per decomposition, number of classes per microservice and the domain modularity.



\section{Conclusion and Future Work}

In this paper, we proposed a hierarchical clustering based approach to decompose a given monolithic application into a set of microservices. The presented approach groups together semantically and structurally similar classes using a modified density-based clustering algorithm and provides as a result a hierarchical structure of the potential microservices as well as outlier classes. We evaluated our approach using different performance metrics and compared it to multiple baselines. The experimental results show that our method achieved better cohesion within the microservices and less interactions between the microservices.

In the future, we plan on developing more fine-grained metrics for evaluating the extracted microservices and comparing them with existing decompositions. We plan as well on investigating the impact of separating utility classes from domain classes and reviewing methods on how to automatically identify them. In addition, we are interested in experimenting with different similarity metrics as well as different types of interactions between classes other than direct method calls.


\bibliographystyle{ACM-Reference-Format}
\bibliography{references}


\end{document}